# Initial experiments concerning quantum information processing in rare-earth-ion doped crystals

M. Nilsson*, L. Levin, N. Ohlsson, T. Christiansson and S. Kröll

Atomic Physics, Lund Institute of Technology, P.O. Box 118, SE-221 00 Lund, Sweden

## Abstract

**In this paper initial experiments towards constructing simple quantum gates in a solid state material are presented. Instead of using specially tailored materials, the aim is to select a subset of randomly distributed ions in the material, which have the interaction necessary to control each other and therefore can be used to do quantum logic operations. The experimental results demonstrate that part of an inhomogeneously broadened absorption line can be selected as a qubit and that a subset of ions in the material can control the resonance frequency of other ions. This opens the way for the construction of quantum gates in rare-earth-ion doped crystals.**

## 1. Introduction

There is an ongoing search for robust systems that can act as hardware for quantum information processing. Ion traps [1], NMR [2], cavity QED [3] and other systems have shown very promising results when it comes to demonstrating control of systems at the quantum level. However, these systems have all proven to be nontrivial to scale to larger number of qubits. Quantum gates in solid-state systems (which are thought to be more scaleable) are yet to be demonstrated. Technical developments have lately made it possible to build tailored materials, with control over the constituents down to the atomic/molecular level. However, there also exist non-tailored complex quantum systems that can be controlled and manipulated. The rare-earth-ion doped crystals constitute one example of such systems and are in our opinion realistic candidates for quantum computer hardware. There have been several proposals concerning the implementation of quantum gates in these materials [4,5,6].

The rare-earth-ion doped crystals have several very attractive features when it comes to implementing quantum gates. Because of the existence of a partially filled inner shell (4f) that is shielded from the environment by outer electrons, they have optical transitions with very narrow line-widths. For Eu doped into $Y_2SiO_5$, some transitions have homogeneous line-widths of less than 1 kHz [7]. The narrow line-width makes it possible to coherently manipulate the ions with long sequences of laser pulses. Most of the rare-earths have a hyperfine splitting of the ground state, due to nuclear quadrupole interaction. The relaxation between different hyperfine levels is very slow and lifetimes can be up to several hours or even days [8]. Measurements of the dephasing time between the hyperfine levels are lacking for many materials, but it is at least of the order of milliseconds for some combinations of dopants and hosts. When doped into inorganic host crystals, the differently located ions experience shifts of their optical absorption frequencies because of imperfections in the crystal host lattice. Because of the differences between different positions in the lattice, the shifts will be different for different ions, creating an inhomogeneous broadening of the optical transition. The broadening can be several GHz, making it possible to address more than $10^6$ different spectrally distinct channels.

* e-mail: mattias.nilsson@fysik.lth.se

In an earlier paper, a scheme for implementing quantum gates in rare-earth-ion doped inorganic crystals has been suggested [6]. The two-level systems constituting the qubits in this scheme are two of the hyperfine levels of the ground state of the ions (figure 1). Different qubits can be addressed by tuning the laser to different optical frequencies within the inhomogeneously broadened optical transition. In the scheme, each qubit will consist of a large number of ions since each frequency channel in the crystal is occupied by many ions. Using optical pumping on the optical transition, all of the ions in one frequency channel can be pumped into the same hyperfine state, creating a well-defined initial state for the qubit. Single qubit operations such as population transfer between the qubit levels can be performed using resonant Raman transitions.

When excited on the optical transition, the ions experience a change in their permanent dipole moments [9]. The change in dipole moment for an ion changes the electric field emanating from it. This change in electric field affects neighbouring ions in the crystal lattice, by changing their optical absorption frequency [10,11,12]. If the excitation of the ions in one qubit changes the absorption frequency for the ions in another qubit sufficiently, the interaction can be used for designing quantum gates. It is worth noting that the frequency shifts for the different ions in the qubit to be controlled do not need to be the same for all ions. It is sufficient that the frequency shifts for all ions are large enough so that laser pulses that are originally in resonance with the ions, after the shift have a negligible effect on the ions, i.e. do not excite them on the optical transition.

The magnitude of the interaction between the ions depends on the distance between the ions. It is advantageous to use the frequency domain for addressing the qubits, since the interaction mechanism has a spatial dependence. Strongly interacting ions would be positioned so close in space, that it would be difficult to spatially address only one of them, without affecting the other.

Since the ions in any arbitrarily chosen frequency interval are randomly positioned in the crystal, only some of the ions in two such intervals are located close enough to be able to control each other. It is therefore necessary to select only the strongly interacting ions for the qubits within these two frequency intervals. The selection procedure has been described in [6] and is briefly recapitulated below.

First all the ions in the two qubits that should be able to control each other are prepared in one of their hyperfine levels. Then all the ions of the controlling qubit are excited on the optical transition. Now the resonance frequencies of the ions in the other qubit, the target qubit, will shift by different amounts, depending on their distance and position relative to the ions in the crystal that were excited. This will cause the originally sharp frequency channel to spread out in frequency. The frequency of some of the ions in the target qubit will have shifted by large amounts and these ions are suitable to use for the qubit. The other ions will still have an absorption frequency close to their original one. It is therefore possible to remove these unwanted ions from the target qubit by means of laser pulses resonant with the original absorption frequency of the ions. In this way, the ions can be transferred to a third hyperfine level of the ground state, which we call the auxiliary level (figure 1). After the removal of the unwanted ions, the ions in the controlling qubit can be brought back to their original state by means of a π-pulse. After this we have a situation where the excitation of the ions in the control qubit will shift all the ions in the target qubit out of their original resonance. By applying the above procedure a second time, but now with the former target qubit as the control qubit, the qubits will be able to control each other mutually. This scheme for distilling out the strongly interacting ions for the qubits, can be generalised to several qubits, making many-qubit gates possible.

In the proposed scheme, the qubits have the shape of narrow spectral peaks (e.g. 1 MHz) within wider (e.g. 10 MHz) "wells" that have previously been emptied by means of optical pumping. This is because it is necessary for all ions belonging to a qubit to experience the same pulse area when a laser pulse is applied to the system. Pulses where all ions that interact with the radiation see the same pulse area are referred to as hard optical pulses in [13]. With the proposed structure it is possible to apply pulses that are spectrally wide so that all ions within a qubit feel the same pulse area, without appreciably exciting ions outside the qubit structure.

Initially the qubits must be prepared by selecting ions absorbing within a particular frequency interval and then selecting the strongly interacting ions before performing the gate operations. In this paper we have experimentally shown and demonstrated selection of ions in a frequency interval and preparation of the ions in a particular hyperfine state in $Eu^{3+}$:$Y_2SiO_5$. We have also characterised the size of the shift in absorption frequency occurring due to dipole-dipole interaction in rare-earth-ion doped materials in theory, and experimentally in $Tm^{3+}$:YAG.

## 2. Preparation of a quantum bit

In this section it is described how a group of ions, absorbing at a particular frequency, is isolated and prepared in a specific hyperfine state. The term qubit will be used for the prepared structure, since it is possible to apply one-bit quantum gates to it, although only the sub-set of ions that interact sufficiently with ions in other qubits will be selected for the final qubit.
The first part of the section describes the material used and the process of spectral holeburning by optical pumping of ions between hyperfine levels. Next it is described how this process can be used for forming the qubit structure in the frequency domain and how this has been achieved experimentally. Finally, the results are compared to simulations and from the simulations conclusions can been drawn as to what materials would be suitable for further experimental work.

The material $Eu^{3+}$:$Y_2SiO_5$ was chosen for the initial qubit preparation experiments. Optical pumping, spectral hole-burning and dephasing times in this material have been studied extensively [7,8,14] and the separations of the hyperfine levels of the $Eu^{3+}$ ions (see figure 1) were suitable for the light source and frequency modulators available. The Eu nucleus has spin $I={}^5/_2$, and the ground and excited states are both split by the nuclear quadrupole interaction into three Kramer's doublets ($I_z=\pm^1/_2\ \pm^3/_2$ and $\pm^5/_2$), which in the following are referred to as hyperfine levels [15]. Analysis is complicated by the fact that the Eu ions can be located at two inequivalent sites (I and II) in the host, with different absorption frequencies, and that there are two Eu-isotopes ($^{151}Eu$ and $^{153}Eu$) of near equal abundance, with different quadrupole moments and hence different splittings of the energy levels. The optical line at 580.05 nm, arising from the $^7F_0$ - $^5D_0$ transition for site II Eu ions, was selected. This transition has a radiative lifetime of 1.7 ms and a dephasing time of up to 0.8 ms at 6 K [16]. The life-time of the ground state hyperfine levels (the spectral hole life-time) is of the order of hours or days at liquid helium temperatures and, although the spin state dephasing time is not known, it can be expected to be longer than the optical dephasing time (which would be an absolute lower limit for the spin dephasing time).
The level structure and some of the separations are shown in figure 1. (Level splittings taken from [14] and [16]). The levels can be labelled either according to quantum number for the nuclear spin orientation, or according to their intended use as qubit states |0>, |1> and |aux>.

If a narrow-band source excites part of the inhomogeneous absorption profile of the material, at frequency $\nu_0$, the ions will be redistributed among the hyperfine levels, leading to decreased absorption (holes) at some frequencies, including $\nu_0$, and increased absorption (anti-holes) at other frequencies [17] (figure 2). It is important to note that, due to the inhomogeneous broadening of the optical transition, the frequency $\nu_0$ might for some ions connect the lowest hyperfine levels of the ground and excited states, while for other ions the frequency $\nu_0$ may connect e.g. the second sublevel of the ground state with the highest level in the excited state etc. Any ion in a particular hyperfine level will contribute to absorption at three different frequencies within the inhomogeneous absorption profile. When burning a hole, i.e. optically pumping away ions at a particular frequency, the result will be that in total 6 side-holes and 42 anti-holes are created, if all transitions are allowed. If two isotopes are present, the number of side-holes and anti-holes will double. During optical pumping, ions relax back to the ground state through a complex route of transitions and it is generally assumed that they are equally probable to end up in any of the three hyperfine levels of the ground state. The relative transition probabilities between different hyperfine levels of the ground and excited states can be calculated from the relative heights of holes and anti-holes when scanning over the created spectral structure. If they are not equal, all pulse areas (pulse power or duration) can be adjusted to fit the actual transition probabilities, or a sequence of pulses with a pulse area of $2\pi$, separated by more than $T_1$, can be applied to pump away ions with a lower/higher oscillator strength [13].

If frequency is to be used for addressing the qubits in the material, one must make sure that all the ions at a particular frequency belong to the same hyperfine level and isotope, by optically pumping away all other ions. If we burn repeatedly at three frequencies, with a separation equal to the separations of the ground state hyperfine levels, only ions for which these three frequencies fit the transitions from the ground hyperfine levels to one of the hyperfine levels in the optically excited state, will remain. All other ions will be pumped to states where they do not interact with light at the chosen frequencies. The picture is simplified in materials such as $Eu^{3+}:Y_2SiO_5$, where the hyperfine splitting of the excited state is much larger than the splitting of the ground state, since one does not have to consider to which hyperfine level an ion is excited.

The set-up that was used is shown in figure 3. Part of the light from a frequency stabilised CW dye laser (Coherent 699-21), pumped by an Argon-ion laser, was passed through two acousto-optic modulators (AOMs) in series, which off-set the frequency by 120-200 MHz (Beam 1). Another part of the light was passed through a single AOM, giving an offset of 60-100 MHz (Beam 2). The beams were added on a beam-splitter and passed through a $Eu^{3+}$:$Y_2SiO_5$ crystal (10 mm long, 0.1% Eu doping concentration), immersed in liquid helium (~4K). Beam 2 was used to access the transition between the lowest hyperfine level of the ground state (|aux>) and an excited state and beam 1 was used to access transitions from the two higher hyperfine levels (|0> and |1>) and to scan over these levels, in order to investigate the spectral holes that were burnt. An additional mechanical shutter was used to block any light leaking through the AOMs when they were turned off. The light transmitted through the sample was detected and divided by the signal from a reference beam, in order to cancel the effect of fluctuations in laser amplitude and AOM efficiency. A matched pair of electronically amplified photodiodes was used for detection. The laser power and beam diameter at the sample were approximately 5 mW (in either path) and ~1 mm, respectively.

The pumping sequence is shown in figure 4. First, the light frequency is scanned repeatedly over 10 MHz wide regions around states |0> and |1>, putting all ions with an appropriate transition frequency between the optical ground and excited states in the |aux> level and pumping away all others. The two “wells” can then be studied by scanning the frequency over a larger interval and monitoring the transmission. If the wells are completely emptied, there should be no absorption at those frequencies. Second, we pump on a narrow frequency region (~2 MHz, width determined by the laser) around the |aux> state and on the “well” around state |0>, letting ions relax back only into state |1>. If the absorption is now studied, we see that a peak has formed in the middle of the well around state |1>. This peak is comprised of ions that can be pumped as desired between states |0>, |1> and |aux>, or manipulated using e.g. “π” or “π/2” pulses. If, for instance, states |1> and |aux> are simultaneously pumped on, all ions end up in the |0> state. Figure 5 shows an example where a majority of ions have been moved into state |1>.

The results shown in figure 5 contain an amount of seemingly irregular structure in the absorption plot, that one might not expect from the simple conceptual pictures. Some of the structure stems from difficulties in controlling the light frequency and holeburning efficiency and can be avoided, but most of it is caused by the formation of holes and anti-holes, as mentioned earlier. The fact that ions can absorb at three different frequencies and the presence of another isotope with different absorption frequencies turn out to put severe limitations on what frequency intervals can be completely emptied of absorbing ions. With the hyperfine level separations of $^{151}Eu$ and $^{153}Eu$ at site II in $Y_2SiO_5$, it turns out that it is actually impossible to completely empty wells symmetrically around states |0> and |1>. This is a serious problem, since the proposed quantum gate operations requires the ability to selectively excite ions belonging only to one qubit, using pulses with a well-defined pulse area. Thus other ions absorbing on the same or nearby frequencies are unacceptable.

A computer program for simulating optical pumping between hyperfine levels in rare-earth-ion doped crystals was developed and used for testing the qubit preparation scheme on various materials, using the hyperfine level separations and pulse sequences as inputs. The program keeps track of the percentage of ions in each hyperfine state, for different transition energies between the electronic states, and assumes fixed probabilities (usually 1/3) for an excited ion to relax back into each of the different hyperfine levels of the ground state. By assuming suitable frequency and power of the laser pulses, the results from the experiments in $Eu^{3+}:Y_2SiO_5$ could be accurately reproduced in simulations (figure 6). Additional simulations were made, in order to find out what material parameters would be more suitable for isolating qubits in a pure state. The simulations show that, as might be expected, in general it is possible to burn wider "wells" around two particular frequencies in materials with just one isotope and where the hyperfine level separations are large. For instance, in a $Y_2SiO_5$ crystal doped with pure $^{153}Eu$ it would be possible to completely empty two intervals, up to 11.9 MHz wide, 76.4 MHz apart, so that the states $|I_z=\pm{}^1/_2\rangle$ and $|I_z=\pm{}^3/_2\rangle$ could be used as qubit states $|0\rangle$ and $|1\rangle$. However, if the frequency separations are too large AOMs cannot be used for frequency modulation of the laser light, and also the number of qubits that can be accommodated within the absorption profile of the transition decreases.

From simulations, $Eu^{3+}:YAlO_3$ was found to be a suitable material for continued experiments. In this material the two higher lying ground state hyperfine levels, $|I_z=\pm{}^1/_2\rangle$ and $|I_z=\pm{}^3/_2\rangle$, of $^{153}Eu$ can be used as qubit states, since it is possible to isolate e.g. a 2 MHz wide group of ions in 14 MHz wide wells, without interference from ions in other states or from the other isotope (figure 7). Another advantage of using Eu doped $YAlO_3$ is that the permanent dipole moments and hence the interaction strength between ions, is larger in this host [9,12,18].

To summarise we have attempted to pump all ions within a selected frequency interval into a pure qubit state (figure 5). By modelling the optical pumping process it is shown (figure 6) that background signal and remaining absorption within the selected frequency interval is due to coincidences in the hyperfine and isotope separations in the $Eu:Y_2SiO_5$ crystal. However, our simulations show (figure 7) that there are crystals and isotopes where the type of remaining backgrouns signal seen in figure 5 should be absent.

## 3. Dipole-dipole interaction

The interaction between qubits in the presented scheme is assumed to occur as shifts of resonance frequencies, due to dipole-dipole interaction between the ions of one qubit and nearby ions belonging to another qubit. In order to characterise this type of interaction, systematic measurements were made of frequency shift as a function of the density of excited ions in a rare-earth doped material. The measurements employed spectral holeburning in the 793 nm ($^3H_6$-$^3H_4$) transition of $Tm^{3+}$:YAG, since a laser source was available at this wavelength that could be tuned single mode over an interval of up to 50 GHz at a rate of more than 1.5 GHz/μs [19]. It has been suggested that the permanent dipole moment of dopant ions in YAG is small or almost vanishing, due to the high symmetry of the ion sites. However, we assumed that by exciting a sufficiently large number of ions it would be possible to measure the effect of a group of excited ions on the other ions in the material.

A spectral hole was burnt in the inhomogeneously broadened absorption profile, after which a varying amount of ions in another part of the profile was excited (henceforth referred to as the perturber excitation). This caused the absorption frequencies of some of the ions at frequencies close to the spectral hole to shift into the hole and the frequencies of others to shift away from the hole. As a result the hole became broader and shallower and from the change in hole shape, conclusions could be drawn regarding the magnitude of the frequency shifts. In order to further investigate the nature of the frequency shifts, complementary experiments were made. In these experiments the ordering of the perturber excitation and the holeburning pulses was changed and the time before monitoring the hole was varied.

An external cavity diode laser (ECDL), which could be rapidly tuned over several GHz within the absorption line of $Tm^{3+}$:YAG, by applying voltage to an intra-cavity electro-optic crystal was used. This allowed for excitation of a significant fraction of the absorption profile (i.e. of all $Tm^{3+}$ ions in the crystal) within the lifetime of the radiative transition ($T_1 \approx 800$ μs [20]). The sample was 1 mm thick, with a Tm dopant concentration of 0.5%. The beam diameter at the sample was ~50 μm and the optical power during hole burning and excitation was 15 mW. During read-out of the hole, the laser was attenuated by a factor 100. The frequency scan used for the read-out was created by an acousto-optic modulator (AOM), while keeping the laser frequency constant. The set-up and pulse sequence can be seen in figure 8.

It is assumed that the shift of resonance frequency of an ion, caused by a random distribution of nearby ions being excited and changing their dipole moment, can be described by a displacement function $D(\Delta\nu)$, describing how probable it is that the ion transition frequency will be shifted by a certain amount $\Delta\nu$. The most likely shift will be $\Delta\nu=0$, as the contribution from many ions can average out, but a large shift may occur if an ion close to the “probe” ion is excited or if the contribution from several excited ions happen to work in the same direction. The displacement function describes the shape of a narrow frequency channel after instantaneous spectral diffusion [10], due to excitation of nearby ions, and the shape of a spectral hole after excitation will be a convolution of the original hole shape with this displacement function. The same information can be obtained by creating a step in population at a certain frequency, i.e. the edge of a wider spectral hole, and studying how ions diffuse from the edge into the hole. Detailed knowledge of $D(\Delta\nu)$ would give information about the spatial distribution of ions, strength of nearest-neighbour interactions etc. Conversely, by knowing or assuming the spatial distribution of ions and their dipole moments, one can calculate the frequency shifts due to dipole-dipole interaction, from first principles.

The experimental data in figure 9 shows a spectral hole with various degrees of excitation of the ions in the material. The excitation density was set by saturating a 0-300 MHz wide interval of the absorption line by scanning the laser over it at a rate of <1 MHz/μs. Saturating a 10 MHz interval at the centre of the 10 GHz wide profile means exciting around 0.05% of the total number of ions. As a first approximation, all holes fit well to Lorentzian curves, which means that the original hole shape, given by the (possibly power broadened) laser line, has been convoluted with a displacement function that is close to a Lorentzian. From the widening of the holes it was possible to calculate the FWHM of $D(\Delta\nu)$, which is a measure of the average magnitude of the frequency shift of an ion. The data in figures 9 and 10 was obtained with a 20 μs hole burning pulse, a variable length (0-300 μs) massive excitation pulse and a 6.5 MHz/100 μs attenuated read-out scan. The data in the figures is averaged over 40 shots. The results show that a spectral channel broadens by approximately 3 kHz per MHz excited bandwidth, in $Tm^{3+}$:YAG.

In a second set of experiments, the spectral holes were observed a long time (a few ms) after the perturber excitation. The holes were still visible due to the fact that some of the ions decay back to their ground state via a meta-stable state (lifetime ~10ms) giving rise to a semi-permanent holeburning mechanism. It was expected that the broadened holes would gradually regain their original width as the excited perturber ion population relaxed back to the original state. However, even though attempts were made, it was not possible to verify this effect experimentally. A possible reason for this could be that part of the excited perturber ion population also decays to the meta-stable state, meaning that it takes a relatively long time before the bulk material surrounding the ions contributing to the spectral hole reaches its original state.

A third experiment was performed to further check the nature of the observed frequency shifts. In this experiment, the crystal was first exposed to the broad-band excitation, shortly after which a spectral hole was burnt. In this case, the spectral hole is expected to be narrow if monitored shortly after its creation, i.e. when most of the massively excited ions are still in the excited state. As these ions change their state, the electric field in the crystal is expected to change, causing the hole to widen. This behaviour could be observed experimentally.

The experiments described above clearly show that ion-ion interaction exist. The shape and scaling of the deduced displacement function agrees with what would be expected from randomly located interacting dipoles (see next paragraph). However, the magnitude of the shifts is surprisingly large when considering the structure of the host material. The third of the experiments described above supports the assumption that the interaction is of the dipole type. In contrast, the second experiment does not exhibit the expected behaviour and calls for further investigation, since it can not be ruled out that the ion-ion interaction in Tm:YAG is partly of other origin than the assumed dipole-dipole interaction.

A computer program was created to simulate frequency shifting by dipole-dipole interaction, using a simple theoretical model. The model assumes a completely random distribution of rare-earth ions among the possible sites in the host. The energy shift associated with the interaction between two electrical dipoles is:

$$\Delta E = \left(\frac{\varepsilon + 2}{3}\right) \frac{|\mu|^2}{4\pi\varepsilon\varepsilon_0 r^3}\left(\hat{\mu}_1 \cdot \hat{\mu}_2 - 3(\hat{\mu}_1 \cdot \hat{\boldsymbol{r}})(\hat{\boldsymbol{r}} \cdot \hat{\mu}_2)\right) \tag{1}$$

where $\varepsilon$ is the dielectric constant of the material and $r$ is the distance between the ions. $|\mu|$ and $\hat{\mu}_{1,2}$ are the magnitude and the direction of $\mu_1$ or $\mu_2$, which are the differences in dipole moment in vacuum, between the ground and excited states of ions 1 and 2 respectively. Electric dipoles, corresponding to excited ions, are placed at random locations within a fixed excitation volume and their effect on a central “probe” ion is calculated. The procedure is repeated for a large number of probe ions, rendering a statistical distribution of frequency shifts. This distribution is equivalent to the displacement function $D(\Delta\nu)$, introduced above, and has a width which depends on the amount of excitation and on material parameters such as $\mu$. The simulations show that, when the density of excited ions is low (which would be the case for all experimental situations relevant for the present work) the displacement function is basically a Lorentzian and the width of the broadening is proportional to the density of excited ions. This is consistent with the experimental results and with the discussion of a similar situation (dipolar broadening in magnetically diluted materials) in [21]. As a model system we have used $Eu^{3+}$:$YAlO_3$ which was identified as a suitable candidate in the qubit preparation section of this paper (figure 11).

From the results of the simulations it is possible to predict how many ions would be shifted in frequency by more than a certain amount, at the excitation of varying amounts of other ions. Generally, these ions are the ones that can be entangled with ions at another frequency, by putting the control ions in a superposition between ground and excited state and then attempting to excite ions at the original resonance frequency. The minimum amount that the ions need to shift is determined by the line-width of the laser, the duration of the control pulses and the width of the spectral structures used as qubits. In the case of $Eu^{3+}$:$YAlO_3$ and a stabilised dye laser (line-width ~1 MHz), we are targeting on selecting ions with a frequency shift $|\nu|>5$ MHz, for the qubits (figure 12).

To summarise this section ion-ion interaction is demonstrated in Tm doped YAG (figure 9). Generally the type of interactions demonstrated in figure 9 is attributed to dipole-dipole interactions between excited ions [10-12]. While some of our experiments supports the viewpoint that the observed interaction is of the dipole-dipole type the results are still inconclusive. Based on relatively straight forward simulations it is then shown that if two qubits are selected from the ions within two 1 MHz wide frequency intervals in Eu:$YAlO_3$, around 0.3% of the ions in each interval can be entangled with each other (the remaining 99.7% of the ions would be removed to the auxiliary state as described in the introduction [6]). 0.3% of the ions in a given 1 MHz frequency interval is by far a sufficient number for demonstrating a CONTROL-NOT gate with a good signal-to-noise ratio.

## 4. Summary

Several basic steps for making qubits in order to create quantum entanglement and quantum gates in rare-earth-ion doped crystals. have been carried out, such as isolating ions with a particular optical transition frequency and pumping them into a pure qubit state. It is described how anti-holes and the presence of several isotopes influence the possibility to perform the necessary qubit preparation by optical pumping.
The nature of dipole-dipole interactions in these rare-earth doped materials has also been studied and a relatively simple numerical model, by which it is possible to estimate the percentage of ions that can be entangled for a given situation has been developed.
In conclusion we feel that the results indicate that rare-earth-ion doped crystals are attractive candidates for future experiments concerning entanglement and quantum information processing.

This work was supported by the Swedish Natural Science Research Council, the Swedish Research Council for Engineering Sciences, the Crafoord Foundation and the European Community Information Society Technologies Program under contract IST-2000-30064.

*Figure 1. Hyperfine levels and separations in states $^7F_0$ and $^5D_0$ of site II $^{151}Eu$ ions in $Eu^{3+}:Y_2SiO_5$. Two of the hyperfine levels of the $^7F_0$ ground state are used as qubit states and the optical transition is intended to act as a switching mechanism for turning the interaction between different qubits on and off. The transition is inhomogeneously broadened by ~10 GHz, allowing for adressing of qubits in the frequency domain.*

*Figure 2. Spectral holeburning in the inhomogeneously broadened absorption profile of $Eu^{3+}:Y_2SiO_5$, by optical pumping between hyperfine levels of the optical ground state. The right hand side of the figure illustrates how a given frequency, $\nu_0$, can cause transitions from different hyperfine levels in the ground state for different ions in the material. The population of some hyperfine levels will be depleted, causing an almost permanent decrease in absorption at frequency $\nu_0$, while the increase in population of other levels will create increased absorption at other frequencies.*

*Figure 3. Set-up used for preparation and manipulation of a qubit in the hyperfine states of $Eu^{3+}:Y_2SiO_5$. The CW light from the dye laser is passed through acousto-optic modulators, which modulates the light and offsets the frequency in order to match transitions from different hyperfine levels.*

*Figure 4. Pulse sequence for isolating the ions comprising a qubit in a pure hyperfine state. The first pulse empties two spectral intervals around qubit states |0> and |1>. The second and third pulses transfer population from the auxilliary state |aux> into state |1>. The population in the qubit states can be monitored by observing the absorption of the last pulse in the sequence.*

*Figure 5. Experimental data from the preparation of a qubit in $Eu^{3+}:Y_2SiO_5$. The pulse sequence used for pumping was: 1) 60 scans of 2 ms duration over 12 MHz intervals around levels |0> and |1> 2) 300 ms simultaneous pumping on level |0> and |aux> 3) read-out by scanning the frequency over both wells in 200μs. The peak in the right hand part of the figure consists of ions in qubit state |1>.*

*Figure 6. Simulated qubit preparation in $Eu^{3+}:Y_2SiO_5$. Suitable laser line-width and frequency scans have been used in the model in order to reproduce the experimental data in figure 5.*

*Figure 7. Simulated qubit preparation for $^{153}Eu$ in $YAlO_3$. The hyperfine level separations of the ground state are 59.7 MHz and 119.2 MHz. Two spectral intervals (at the left hand side of the figure) have been completely emptied of absorbing ions before ions from the auxilliary state (at the right hand side of the figure) have been pumped back into state |1>. The rest of the structure (side-holes and anti-holes) is a side effect due to the large number of possible transitions between ground and excited state hyperfine levels.*

*Figure 8. a) Set-up for studying instantaneous frequency shifts in $Tm^{3+}:YAG$. Broad band excitation is achieved by rapidly chirping the frequency of an external cavity diode laser (ECDL). b) Pulse sequence and frequency tuning for creation of a spectral hole, perturber excitation and read-out of the hole.*

*Figure 9. Spectral hole in $Tm^{3+}:YAG$ at increasing levels of excitation of dopant ions. Dotted lines (partly hidden under the experimental traces) are numerical fits to a*

*Lorentzian. The data is recorded by means of absorption measurements. The hole can clearly be seen to broaden as the bandwidth of the perturber excitation increases.*

*Figure 10. Experimentally determined frequency shifts in Tm:YAG due to ion-ion interaction. A hole, originally less than 500 kHz wide, broadens to more than 1.4 MHz at the excitation of ions at other frequencies.*

*Figure 11. Simulated broadening of a frequency channel in 0.5 at.% Eu-doped $YAlO_3$ , due to dipole-dipole interaction. The upper part of the figure shows the broadening of an originally completely sharp frequency channel after the excitation of 10 MHz (stars) and 50 MHz (circles) of the inhomogeneous profile. The solid lines are fits to Lorentzian curves.*

*Figure 12. Numerically estimated percentage of ions shifting more than 5 MHz due to the excitation of ions within a given frequency interval in a different part of the inhmogeneous profile. The simulation is performed for 0.5 at. % Eu-doped $YAlO_3$, with an inomogeneous line-width of 4 GHz and assuming excitation of all ions within the given frequency interval.*

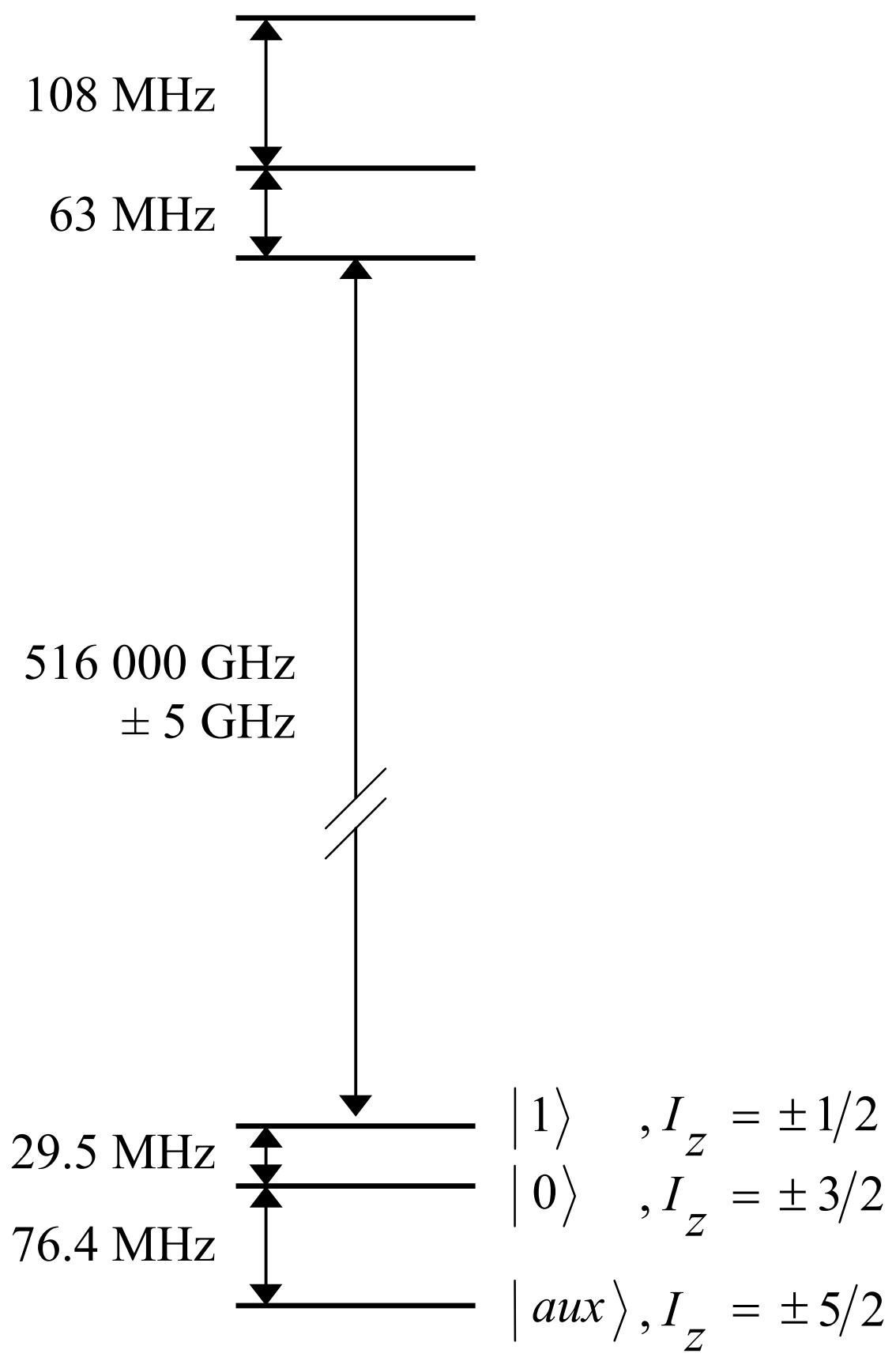

108 MHz
63 MHz
516 000 GHz
± 5 GHz
29.5 MHz
76.4 MHz
$|1\rangle$ , $I_z = \pm 1/2$
$|0\rangle$ , $I_z = \pm 3/2$
$|aux\rangle$, $I_z = \pm 5/2$

Figure 1

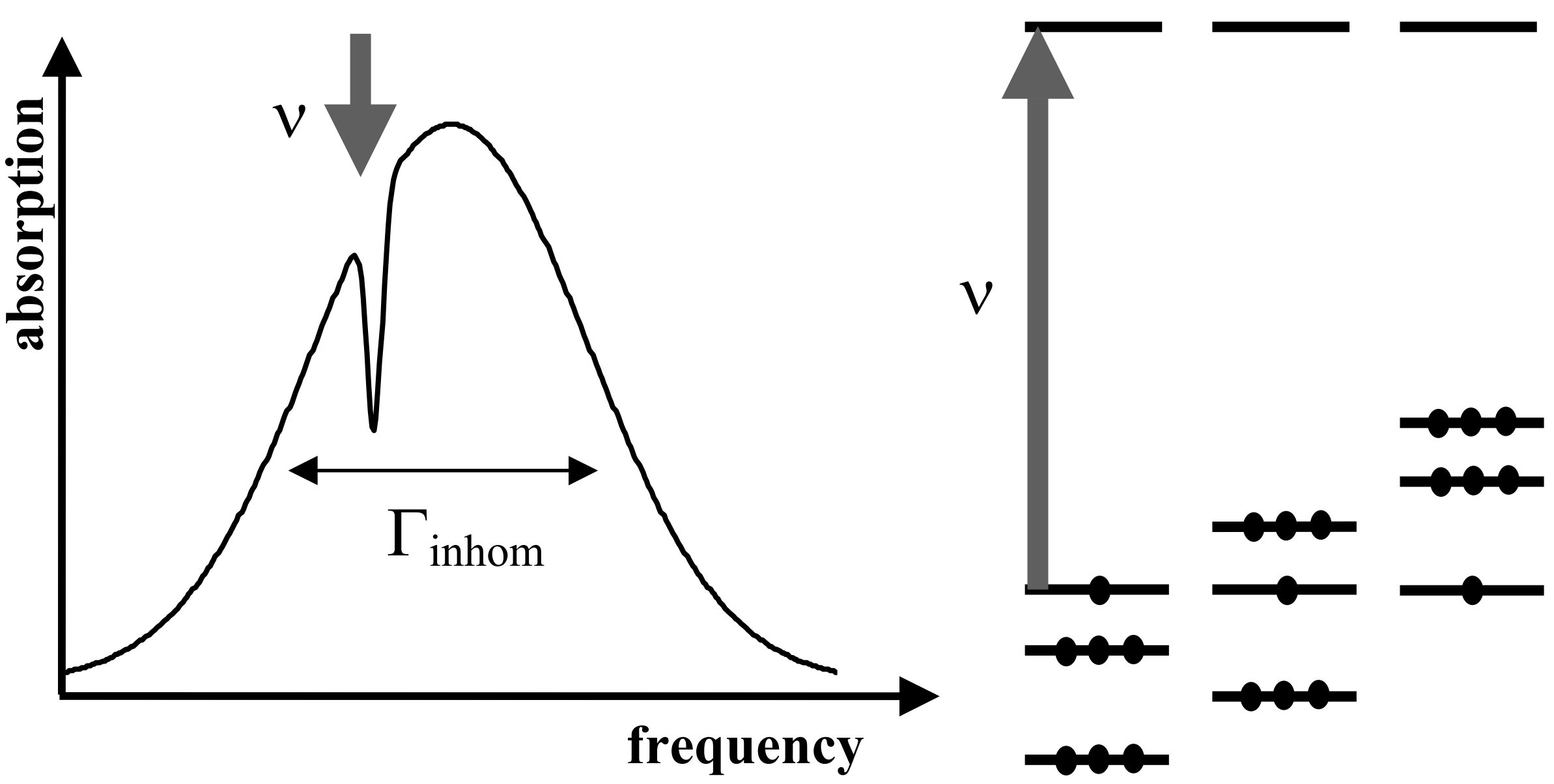

Figure 2

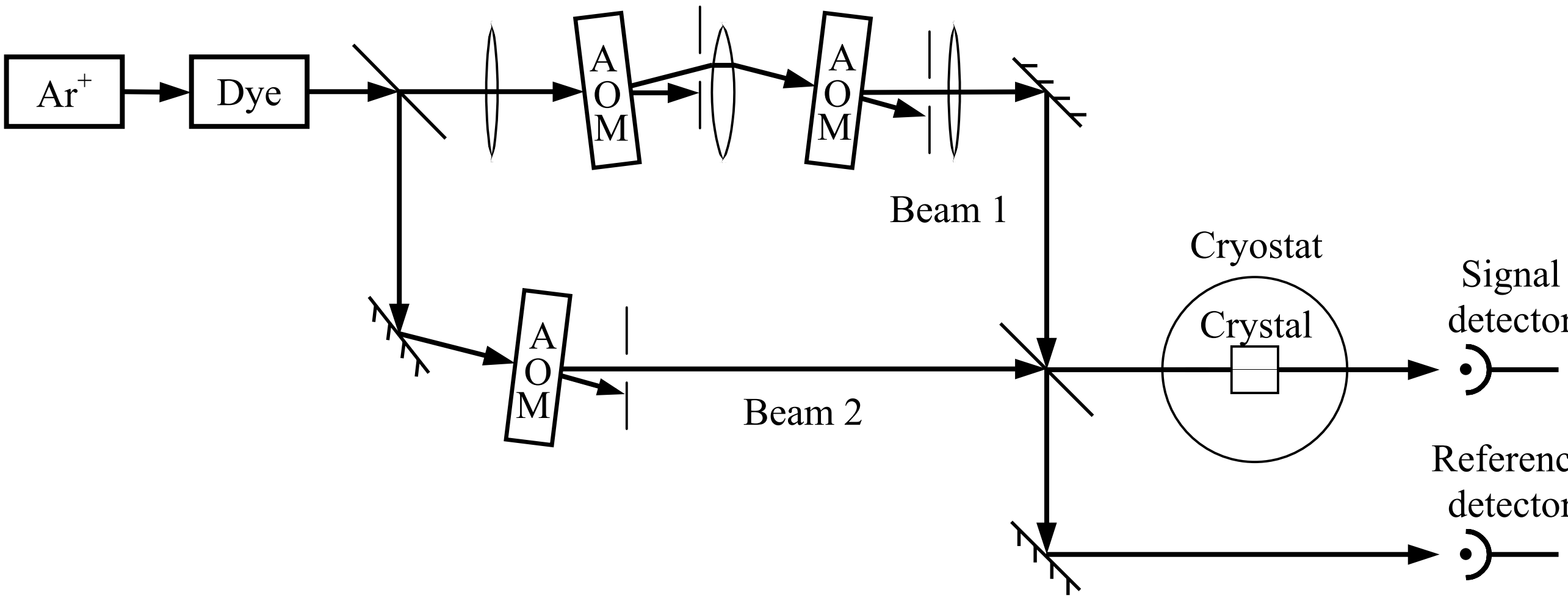

Figure 3

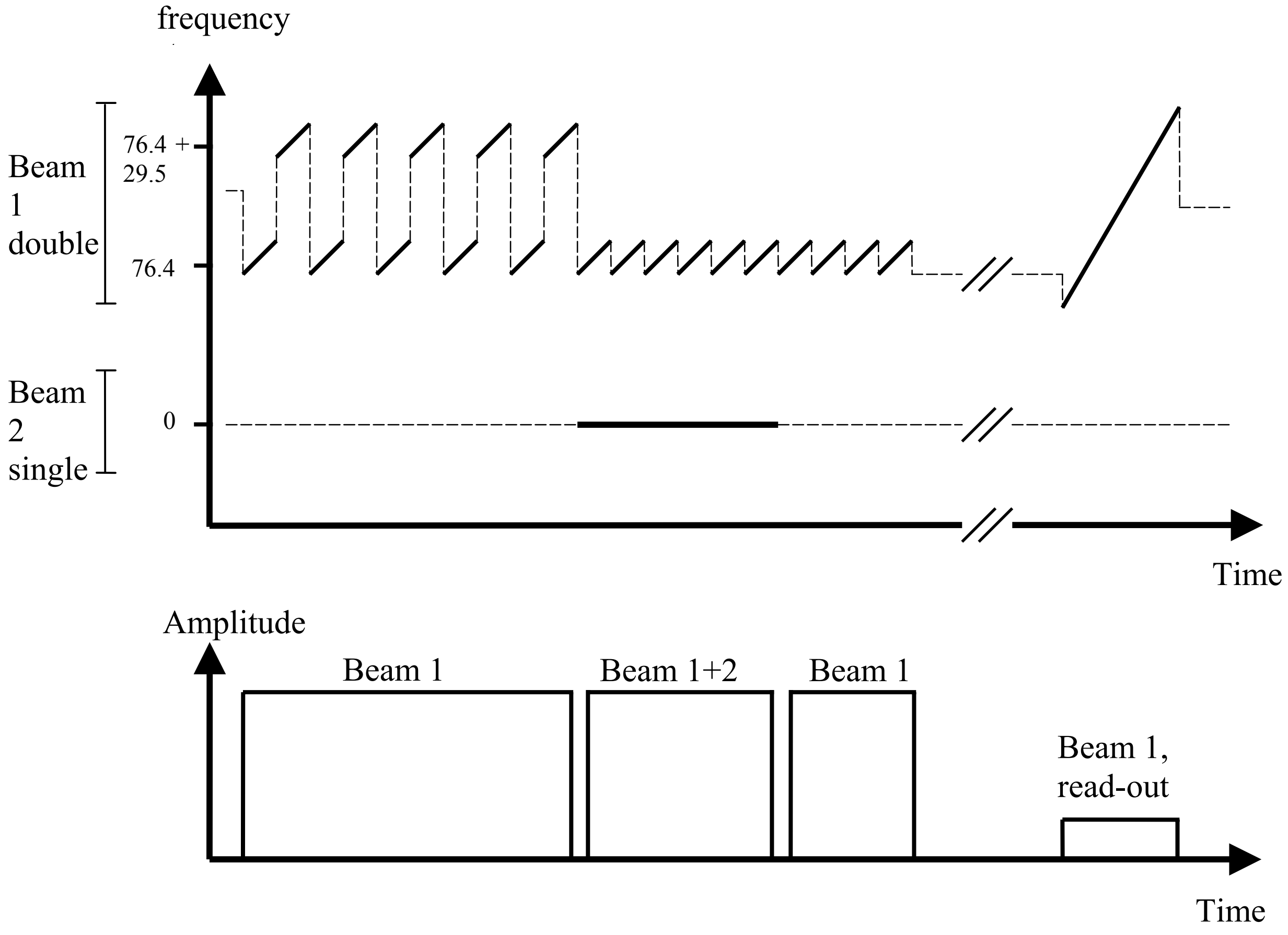

Figure 4

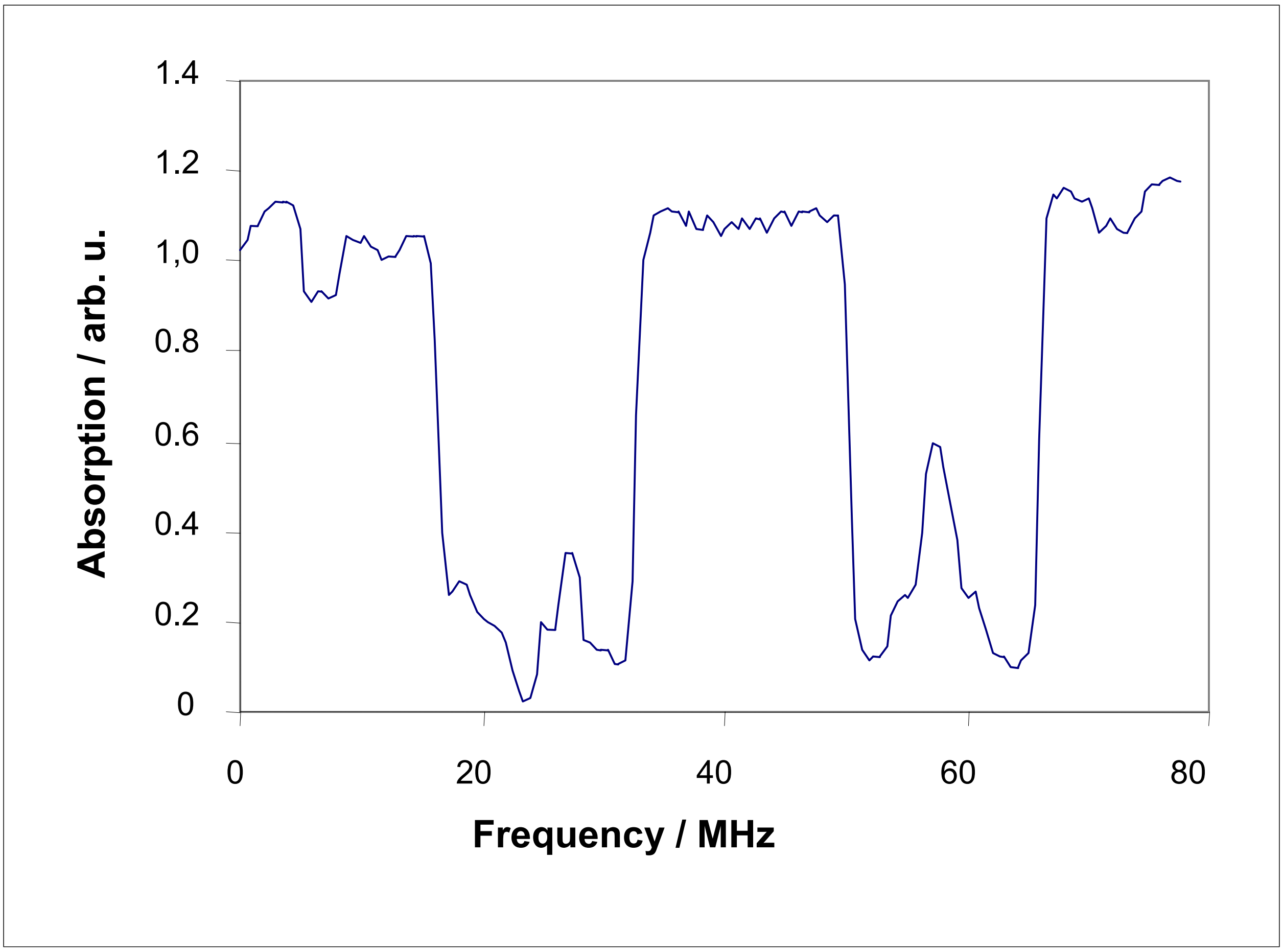

Figure 5

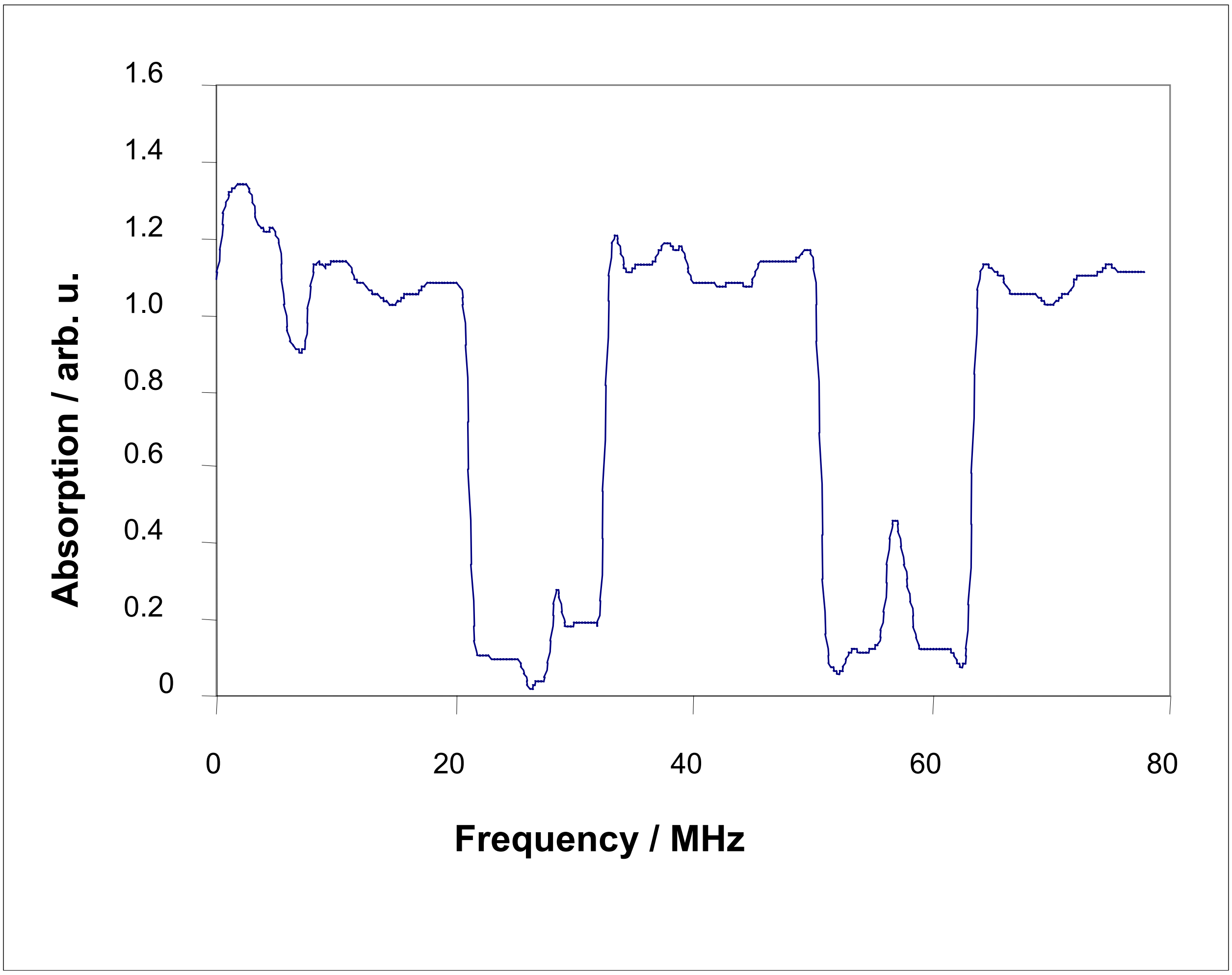

Figure 6

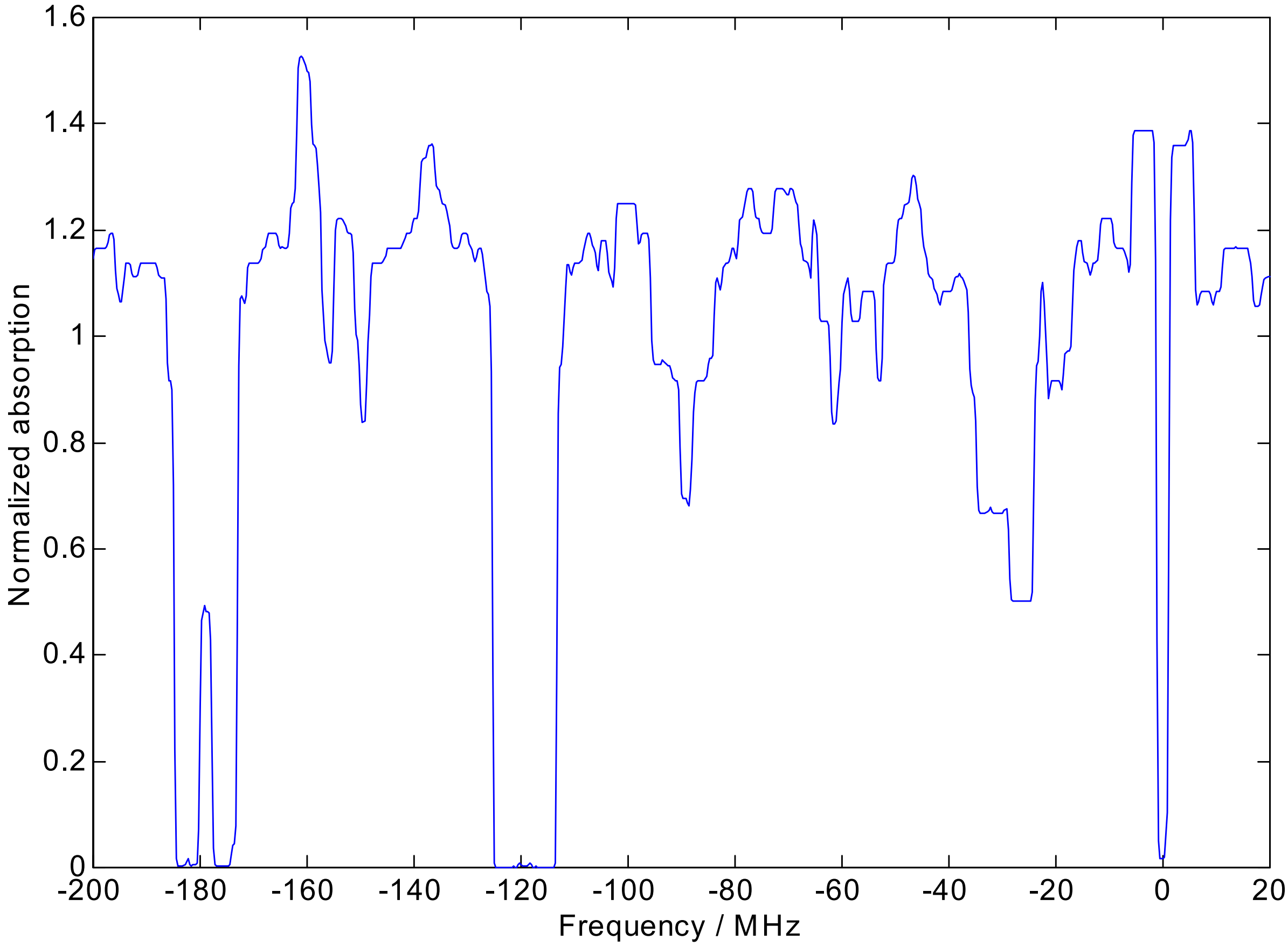

Figure 7

**(a)**

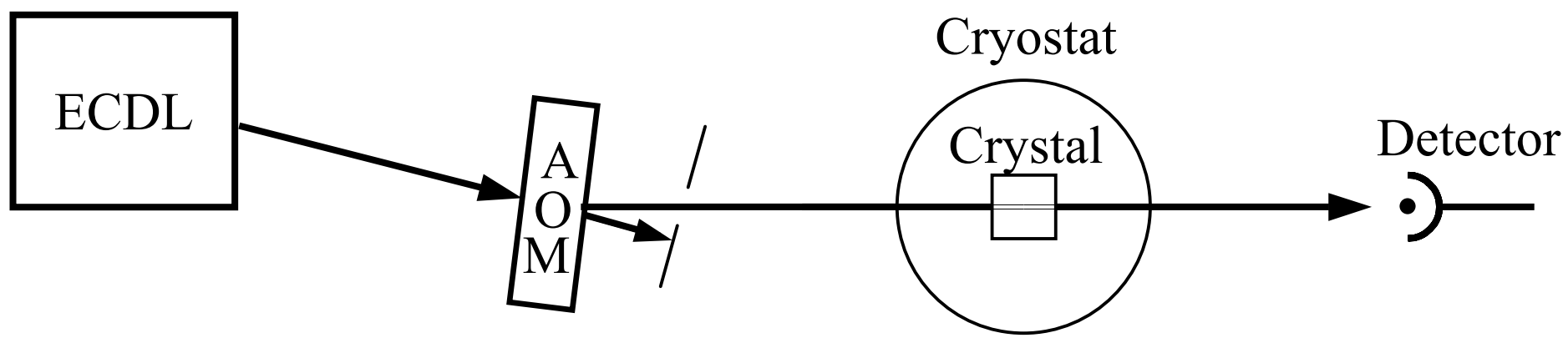

**(b)**

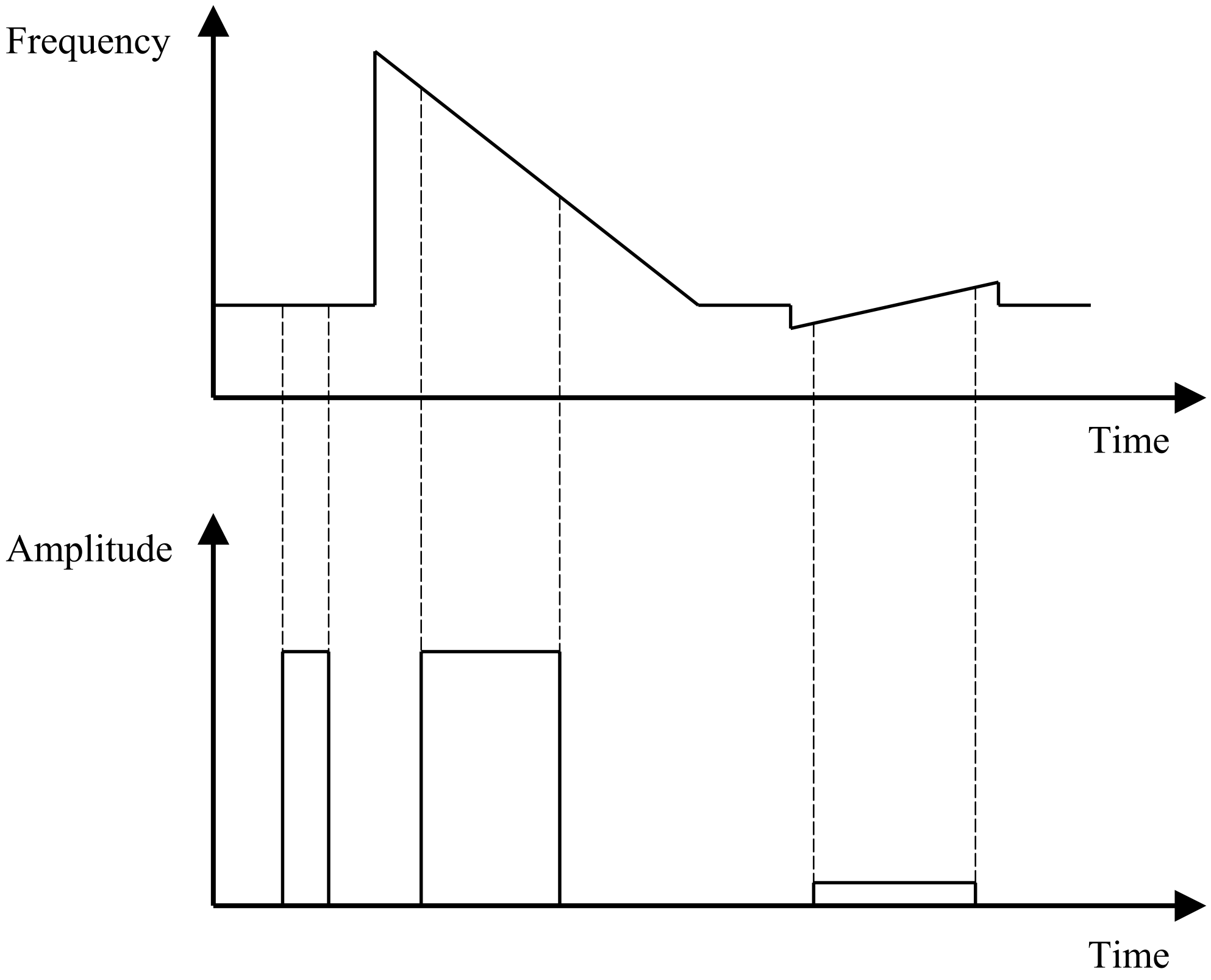

Frequency
Time
Amplitude
Time

# Figure 8

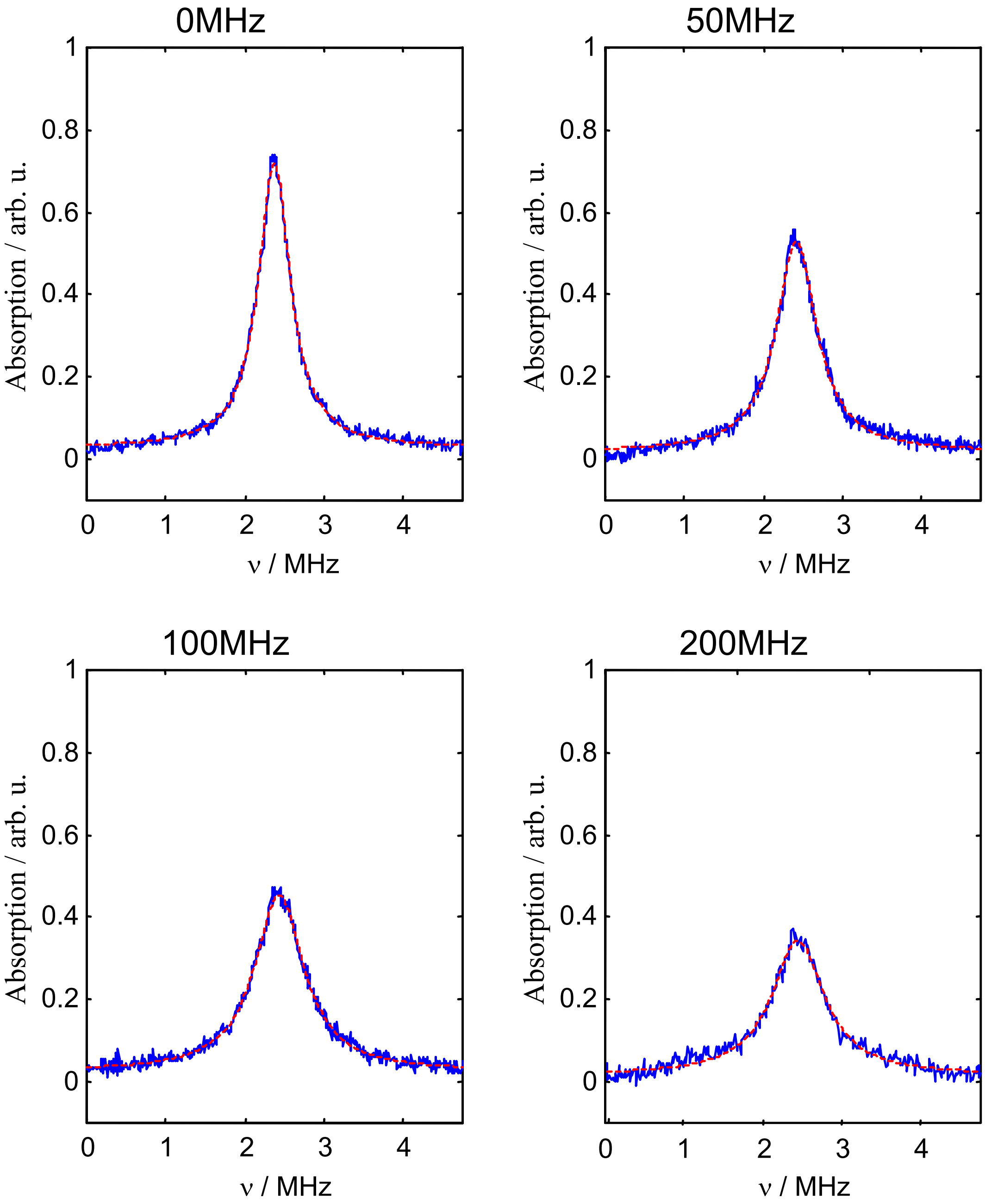

Figure 9

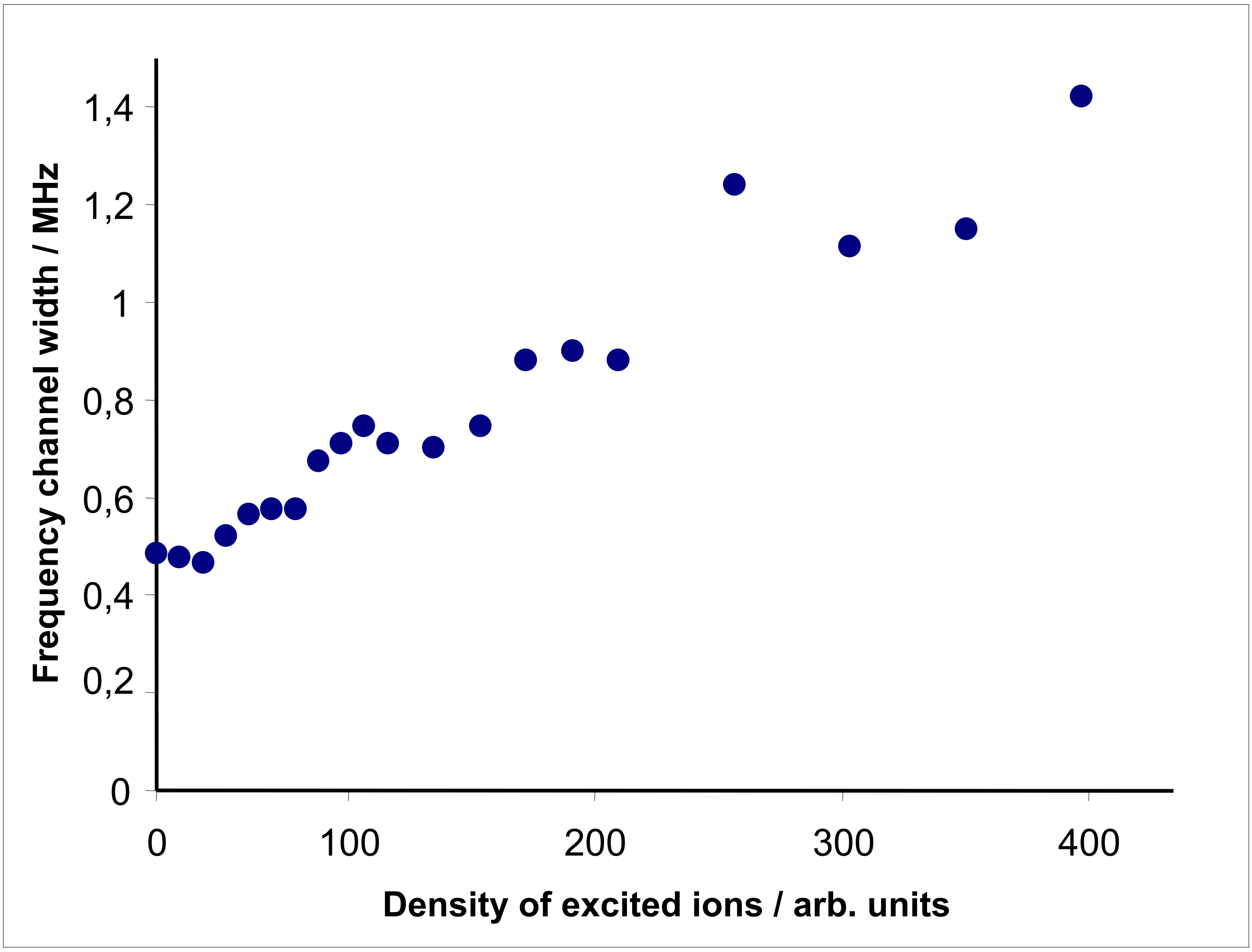

Figure 10

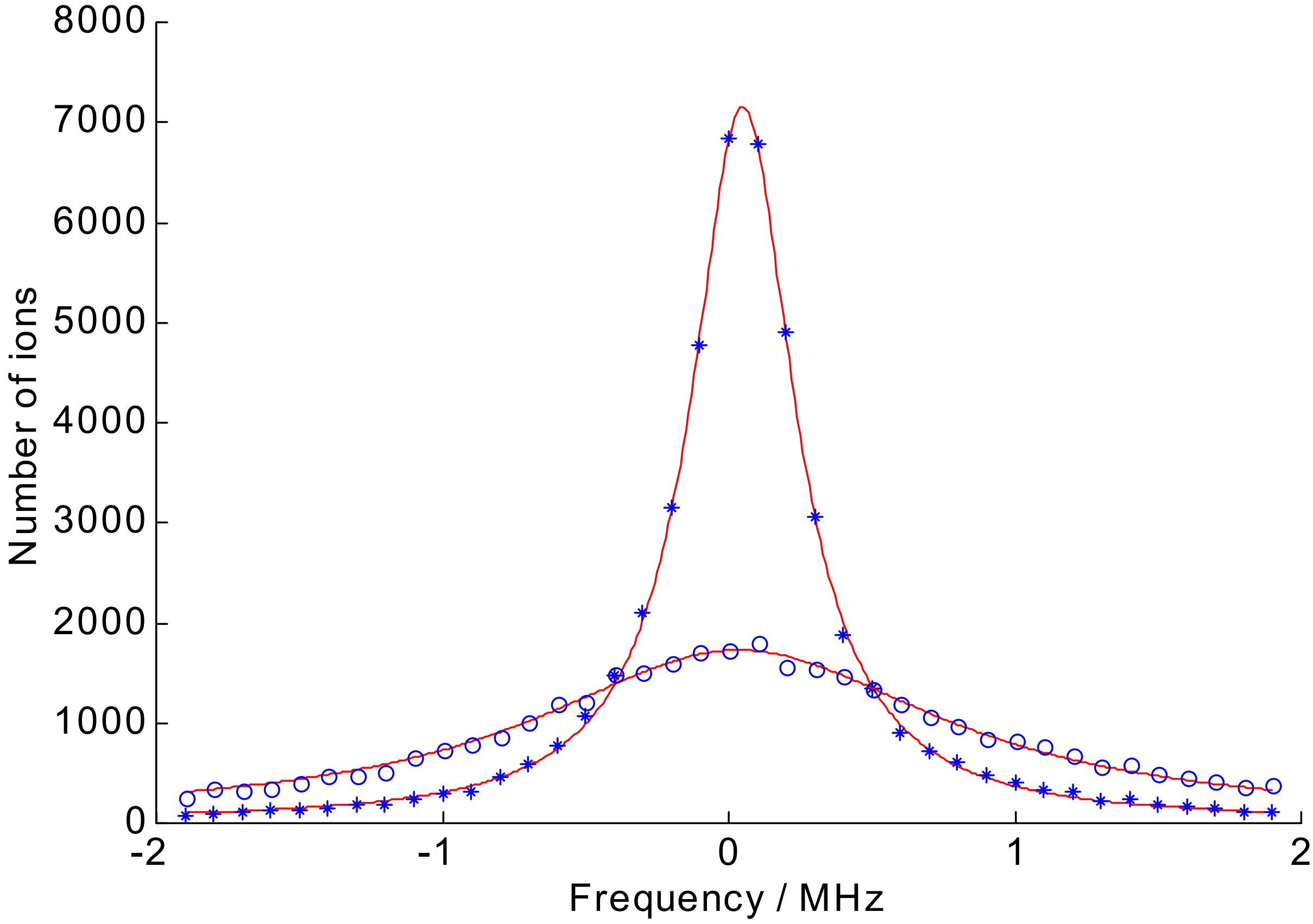

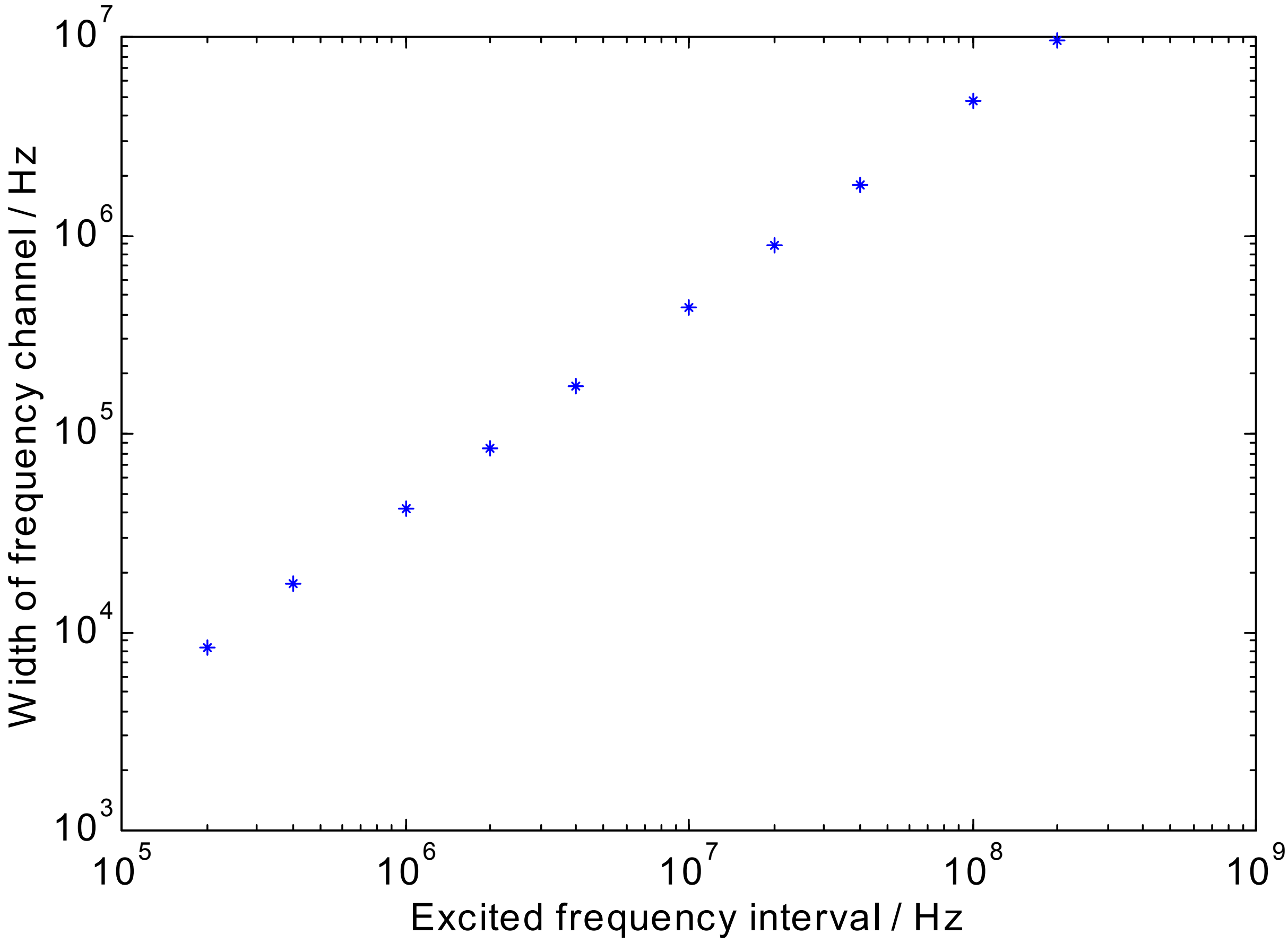

Figure 11

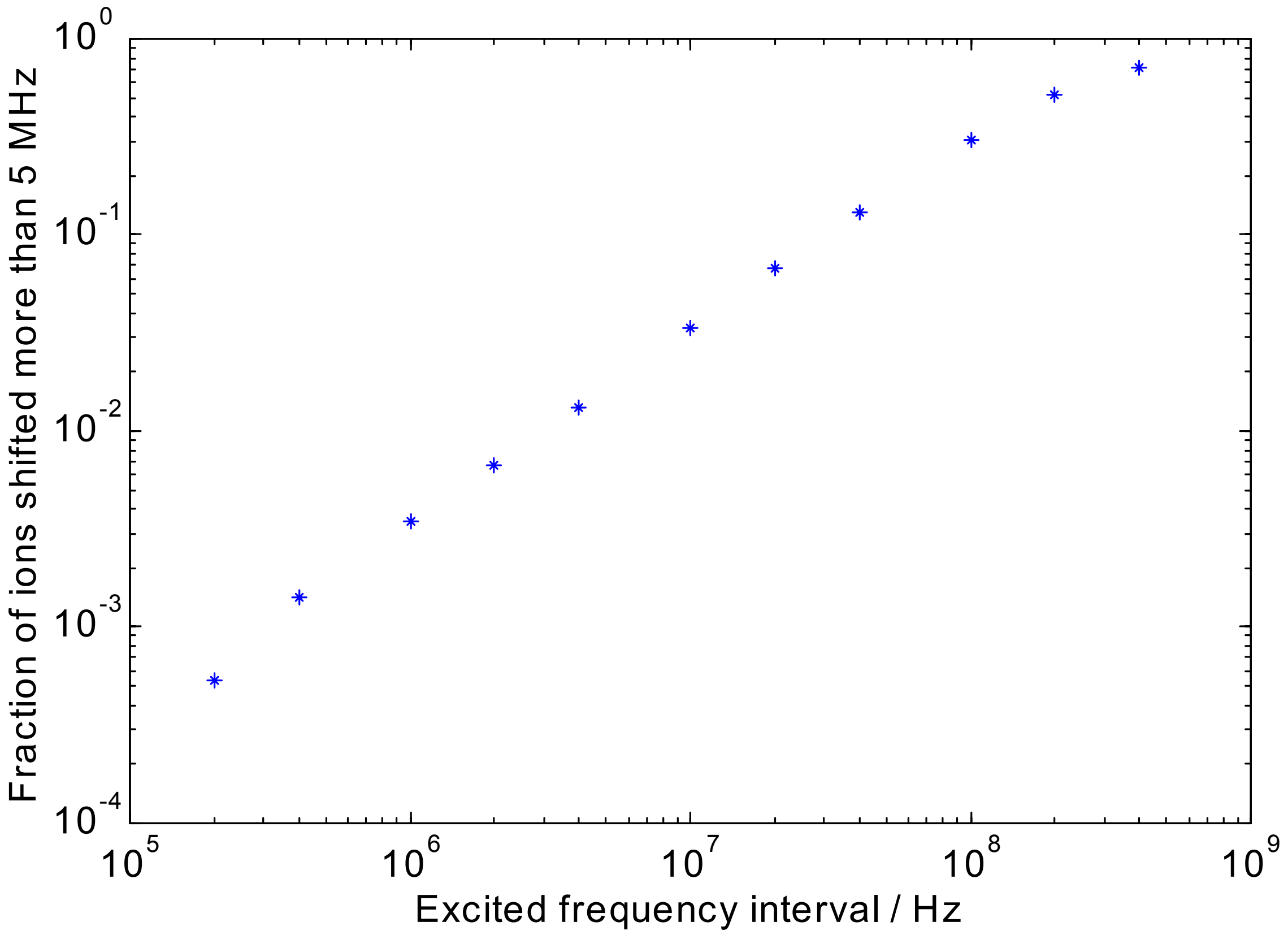

Fraction of ions shifted more than 5 MHz
Excited frequency interval / Hz
10^0
10^-1
10^-2
10^-3
10^-4
10^5
10^6
10^7
10^8
10^9

Figure 12

[1] Sackett, C.A. et al., Nature **404**, 256 (2000)
[2] Gershenfeld, N.A. and Chuang, I.L., Science **275**, 350 (1997)
[3] Turchette, Q.A., Hood, C.J., Lange, W., Mabuchi, H., Kimble, H.J., Phys. Rev. Lett. **75**, 4710 (1995)
[4] Lukin, M.D. and Hemmer, P.R., Phys. Rev. Lett. **84**, 2818 (2000)
[5] Ichimura, K., Opt. Commun. **196**, 119 (2001)
[6] Ohlsson, N., Mohan, K. and Kröll, S., "Quantum Computer Hardware based on rare earth ion doped inorganic crystals", accepted for publication in Opt. Commun.
[7] Equall, R. W., Sun, Y., Cone, R. L. and Macfarlane, R. M., Phys. Rev. Lett. **72**, 2179 (1994)
[8] Könz, F., et al., "Concentration, temperature, and composition dependence of optical dephasing, spectral hole lifetime, and anisotropic absorption in $Eu^{3+}:Y_2SiO_5$", unpublished.
[9] Meixner, A. J., Jefferson, C. M. and Macfarlane, R. M., Phys. Rev. B **46**, 5912 (1992)
[10] Huang, J. Zhang, J. M., Lezama, A. and Mossberg, T. W., Phys. Rev. Lett. **63**, 78 (1989)
[11] Altner, S. B. and Mitsunaga M., Phys. Rev. Lett. **76**, 1747 (1996)
[12] Graf, F. R., Renn, A., Zumofen, G. and Wild, U. P., Phys. Rev. B **58**, 5462 (1998)
[13] Pryde, G. J., Sellars M. J. and Manson N. B., Phys. Rev. Lett. **84**, 1152 (2000)
[14] Yano, R., Mitsunaga, M. and Uesugi, N., J. Opt. Soc. Am. B **9**, 992 (1992)
[15] Macfarlane, R. M. and Shelby, R. M., "Coherent Transient and Holeburning Spectroscopy of Rare Earth Ions in Solids", (Elsevier Science Publishers B. V., 1987)
[16] Yano, R., Mitsunaga, M. and Uesugi, Opt. Lett. **16**, 1884 (1991)
[17] Erickson, L. E., Phys. Rev. B **16**, 4731 (1977)
[18] Graf, F. R., Renn, A., Wild, U. P. and Mitsunaga, M., Phys. Rev. B **55**, 11225 (1997)
[19] Levin, L., "Mode-hop free electro-optically tuned diode laser", accepted for publication in Opt. Lett.
[20] Macfarlane R. M., Opt. Lett. **18** (1993)
[21] Kittel, C. and Abrahams, E., Phys. Rev. **90**, 238 (1953)